\documentclass[11pt,a4paper]{article}

\usepackage{jinstpub}

\usepackage{subfigure}
\usepackage{hyperref}

\begin{document}

\title{Pulse Shape Discrimination in CUPID-Mo using Principal Component Analysis}
\author[a,1]{R.~Huang\note[1]{Corresponding author}}
\author[b]{E.~Armengaud}
\author[c]{C.~Augier}
\author[d]{A.~S.~Barabash}
\author[e]{F.~Bellini}
\author[f]{G.~Benato}
\author[g]{A.~Beno\^{\i}t}
\author[h,i]{M.~Beretta}
\author[j]{L.~Berg\'e}
\author[c]{J.~Billard}
\author[k]{Yu.~A.~Borovlev}
\author[j]{Ch.~Bourgeois}
\author[l]{V.~B.~Brudanin}
\author[g]{P.~Camus}
\author[m]{L.~Cardani}
\author[m]{N.~Casali}
\author[c]{A.~Cazes}
\author[j]{M.~Chapellier}
\author[c]{F.~Charlieux}
\author[n]{M.~de~Combarieu}
\author[m]{I.~Dafinei}
\author[o]{F.~A.~Danevich}
\author[c]{M.~De~Jesus}
\author[a]{T.~Dixon}
\author[j]{L.~Dumoulin}
\author[p]{K.~Eitel}
\author[b]{F.~Ferri}
\author[q]{B.~K.~Fujikawa}
\author[c]{J.~Gascon}
\author[h,i]{L.~Gironi}
\author[j]{A.~Giuliani}
\author[k]{V.~D.~Grigorieva}
\author[b]{M.~Gros}
\author[j]{E.~Guerard}
\author[b]{D.~L.~Helis}
\author[r]{H.~Z.~Huang}
\author[s]{J.~Johnston}
\author[c]{A.~Juillard}
\author[j]{H.~Khalife}
\author[t]{M.~Kleifges}
\author[o]{V.~V.~Kobychev}
\author[a,q]{Yu.~G.~Kolomensky}
\author[d]{S.I.~Konovalov}
\author[j]{P.~Loaiza}
\author[r]{L.~Ma}
\author[k]{E.~P.~Makarov}
\author[j]{P.~de~Marcillac}
\author[j]{R.~Mariam}
\author[a,f,q]{L.~Marini}
\author[j]{S.~Marnieros}
\author[c]{D.~Misiak}
\author[b]{X.-F.~Navick}
\author[b]{C.~Nones}
\author[a]{E.B.~Norman}
\author[j]{E.~Olivieri}
\author[s]{J.~L.~Ouellet}
\author[f,u]{L.~Pagnanini}
\author[n]{P.~Pari}
\author[f,v]{L.~Pattavina}
\author[b]{B.~Paul}
\author[h,i]{M.~Pavan}
\author[w]{H.~Peng}
\author[i]{G.~Pessina}
\author[f]{S.~Pirro}
\author[j]{D.~V.~Poda}
\author[o]{O.~G.~Polischuk}
\author[h,i]{E.~Previtali}
\author[j]{Th.~Redon}
\author[l]{S.~Rozov}
\author[x]{C.~Rusconi}
\author[c]{V.~Sanglard}
\author[j]{J.~A.~Scarpaci}
\author[f]{K.~Sch\"affner}
\author[q,2]{B.~Schmidt\note[2]{Now at: Northwestern University, Evanston, IL 60208, USA}}
\author[r]{Y.~Shen}
\author[k]{V.~N.~Shlegel}
\author[p]{B.~Siebenborn}
\author[a]{V.~Singh}
\author[m]{C.~Tomei}
\author[o]{V.~I.~Tretyak} 
\author[d]{V.~I.~Umatov}
\author[c]{L.~Vagneron}
\author[y]{M.~Vel\'azquez}
\author[t]{M.~Weber}
\author[q]{B.~Welliver}
\author[s]{L.~Winslow}
\author[w]{M.~Xue}
\author[l]{E.~Yakushev}
\author[o]{M.~M.~Zarytskyy}
\author[j]{A.~S.~Zolotarova}

\emailAdd{roger\_huang@berkeley.edu}

\affiliation[a]{University of California, Berkeley, CA 94720, USA}
\affiliation[b]{IRFU, CEA, Universit\'{e} Paris-Saclay, F-91191 Gif-sur-Yvette, France}
\affiliation[c]{Univ Lyon, Universit\'{e} Lyon 1, CNRS/IN2P3, IP2I-Lyon, F-69622, Villeurbanne, France} 
\affiliation[d]{National Research Centre Kurchatov Institute, Institute of Theoretical and Experimental Physics, 117218 Moscow, Russia} 
\affiliation[e]{Dipartimento di Fisica, Sapienza Universit\`a di Roma, P.le Aldo Moro 2, I-00185, Rome, Italy}
\affiliation[f]{INFN, Laboratori Nazionali del Gran Sasso, I-67100 Assergi (AQ), Italy}
\affiliation[g]{CNRS-N\'eel, 38042 Grenoble Cedex 9, France}
\affiliation[h]{Dipartimento di Fisica, Universit\`{a} di Milano-Bicocca, I-20126 Milano, Italy}
\affiliation[i]{INFN, Sezione di Milano-Bicocca, I-20126 Milano, Italy} 
\affiliation[j]{Universit\'e Paris-Saclay, CNRS/IN2P3, IJCLab, 91405 Orsay, France}
\affiliation[k]{Nikolaev Institute of Inorganic Chemistry, 630090 Novosibirsk, Russia}
\affiliation[l]{Laboratory of Nuclear Problems, JINR, 141980 Dubna, Moscow region, Russia}
\affiliation[m]{INFN, Sezione di Roma, P.le Aldo Moro 2, I-00185, Rome, Italy}
\affiliation[n]{IRAMIS, CEA, Universit\'{e} Paris-Saclay, F-91191 Gif-sur-Yvette, France}
\affiliation[o]{Institute for Nuclear Research, 03028 Kyiv, Ukraine} 
\affiliation[p]{Karlsruhe Institute of Technology, Institute for Astroparticle Physics, 76021 Karlsruhe, Germany}
\affiliation[q]{Lawrence Berkeley National Laboratory, Berkeley, CA 94720, USA}
\affiliation[r]{Key Laboratory of Nuclear Physics and Ion-beam Application (MOE), Fudan University, Shanghai 200433, PR China}
\affiliation[s]{Massachusetts Institute of Technology, Cambridge, MA 02139, USA}
\affiliation[t]{Karlsruhe Institute of Technology, Institut f\"{u}r Prozessdatenverarbeitung und Elektronik, 76021 Karlsruhe, Germany} 
\affiliation[u]{Gran Sasso Science Institute, I-67100 L'Aquila, Italy}
\affiliation[v]{Physik Department, Technische Universit\"at M\"unchen, Garching D-85748, Germany}
\affiliation[w]{Department of Modern Physics, University of Science and Technology of China, Hefei 230027, PR China}
\affiliation[x]{Department of Physics and Astronomy, University of South Carolina, Columbia, SC 29208, USA}
\affiliation[y]{Universit\'e Grenoble Alpes, CNRS, Grenoble INP, SIMAP, 38402 Saint Martin d'H\'eres, France}
\abstract{CUPID-Mo is a cryogenic detector array designed to search for neutrinoless double-beta decay ($0\nu\beta\beta$) of $^{100}$Mo. It uses 20 scintillating $^{100}$Mo-enriched Li$_2$MoO$_4$ bolometers instrumented with Ge light detectors to perform active suppression of $\alpha$ backgrounds, drastically reducing the expected background in the $0\nu\beta\beta$ signal region. As a result, pileup events and small detector instabilities that mimic normal signals become non-negligible potential backgrounds. These types of events can in principle be eliminated based on their signal shapes, which are different from those of regular bolometric pulses. We show that a purely data-driven principal component analysis based approach is able to filter out these anomalous events, without the aid of detector response simulations.}

\maketitle

\section{Introduction}

CUPID-Mo is a bolometric experiment searching for neutrinoless double-beta decay ($0\nu\beta\beta$) of $^{100}$Mo, operating 20 scintillating $^{100}$Mo-enriched Li$_2$MoO$_4$ (LMO) crystals complemented with germanium-based light detectors at a base temperature of $\sim$20 mK \cite{CUPIDMoCommissioning}. It has served as a demonstration of the effectiveness of using light signals in combination with heat signals in a scintillating bolometer to discriminate between $\alpha$ events and $\beta/\gamma$ events, which will be key to suppressing backgrounds in the future CUPID experiment \cite{CUPIDInterestGroup:2019inu}. CUPID-Mo has also demonstrated the reproducibility of LMO bolometers with high energy resolution, efficient particle identification capabilities, and high radiopurity \cite{MoDevelopment}, allowing it to achieve the current world-leading limit on the $0\nu\beta\beta$ half-life of $^{100}$Mo \cite{CUPIDMo0v,CUPIDMo0vPaper}.

Physical events in CUPID-Mo are registered when particles deposit energy in a LMO crystal, causing a temperature increase that can be measured with thermistors attached to the crystals. The data acquisition system records a continuous stream of data from each detector, and sufficiently large sudden changes in temperature are triggered as events for processing. Event triggers are produced via an optimum trigger algorithm \cite{CUORE0vLimit}, which evaluates the amplitude of optimum filtered waveforms in the data stream. These triggers are then used to select events for further processing. While a number of basic selection criteria described in \cite{CUPIDMoCommissioning} are applied to eliminate spurious events, most notably including requiring the observation of an amount of light in nearby light detectors consistent with $\beta/\gamma$ interactions, there nonetheless remain a number of undesirable events that should be rejected. Some are non-physical events that pass the trigger threshold of the detectors and are able to slip past these cuts, often as the result of coincident detector noise in a LMO bolometer and an adjacent light detector. Others are pileup events, in which multiple physical events deposit energy in a detector within a short enough time window that our data processing evaluates them as one event. These two classes of events consitute a non-neglible background around the region of interest, as the $^{100}$Mo $0\nu\beta\beta$ Q-value of 3034 keV \cite{Mo100Qvalue} is above most natural $\gamma$ backgrounds and any $\alpha$ backgrounds are rejected using the light detectors. It is thus highly desirable to have a way of rejecting pileup events and other anomalous detector effects emulating normal bolometric pulses.

By looking at the events that pass the basic data quality selection and fall into physically significant energy regions in CUPID-Mo, we can get an idea of the rough pulse shape that most signal-like events should have. However, there are no analogous populations of anomalous pulses that we can learn to discriminate against. The problem can then be stated as an exercise in eliminating anomalous events given a general sample of good events, without the aid of simulation to provide training data with true labels.

\section{Principal Component Analysis}

Principal component analysis (PCA) is a method of rotating data into a new basis with components ranked by how much of the variance in data they explain \cite{PCAOverview}. When decomposed in this way, the leading PCA components are usually able to explain most of the interesting features in the data, while the remaining components tend to explain noise-like features. In the context of calorimetric signal processing, PCA-based methods have been shown to have resolution power superior to standard optimal filtering techniques under certain circumstances \cite{TESPCA,PCAOF}, with the leading PCA components used to estimate the energy of a pulse. Other studies have demonstrated the ability to perform outlier detection using PCA-based methods without having to rely on labeled training data like in supervised learning \cite{PCACalorimeterOutlier,PCAHeavyIonOutlier}.

From a set of $n$-dimensional training data, we can obtain a set of leading components $\mathbf{w}_a$ through the standard PCA decomposition procedure, where each $\mathbf{w}_a$ is an $n$-dimensional vector and the $a=1,2,3...$ are the component rankings. Any $n$-dimensional data $\mathbf{x}$ then has a projection onto each component $\mathbf{w}_a$ given by $q_a = \mathbf{x}\cdot\mathbf{w}_a$. From these projections, we can define the associated \textit{reconstruction error} of $\mathbf{x}$ as
\begin{equation}
\sqrt{\sum\limits_{i=1}^n \left(\mathbf{x}_i - \sum\limits_{a=1}^m q_a\mathbf{w}_{a,i}\right)^2},
\label{eq:RecError}
\end{equation}
where $\mathbf{w}_{a,i}$ is index $i$ of component $\mathbf{w}_a$, and where $m$ can be chosen as desired but will probably be very small since only the leading terms are significant. This metric measures how well one can reconstruct the data $\mathbf{x}$ using only the leading $m$ PCA-derived components. This will result in a large reconstruction error if $\mathbf{x}$ is poorly described using only the leading $m$ PCA components, generally indicating it has some substantial feature difference compared to the training data from which the PCA components were derived. Using a squared metric for the reconstruction error instead of an absolute value emphasizes large localized feature differences over normal noise spread out over many points. We can then attempt to identify anomalous data by looking for large reconstruction errors.

\section{Application in CUPID-Mo}

In CUPID-Mo, waveforms are digitized at 500 samples per second, and an event is taken to be a 3 second window with 1 second before the trigger and 2 seconds after it \cite{CUPIDMoCommissioning}. This can be treated as 1500-dimensional data, with each point simply being the voltage reading across the thermistor at a given time. The typical pulsed response of a LMO crystal to an energy deposit varies from crystal to crystal, depending on factors such as minor variations in the crystal support structures and details of the corresponding electronics and thermistors. To account for the channel-dependence of this pulsed response, we treat each of the LMO channels separately in the PCA procedure. Since the goal is to recognize anomalous pulses by their large reconstruction error, we should try to build PCA components that mostly encapsulate the signal-like information of typical pulses. Pulses that do not follow this signal-like template are then poorly described by the leading PCA components, as desired. However, we also do not want the discriminator to acquire a strong energy dependence. Most of our training and testing must naturally be done in the low-energy $\beta/\gamma$ regions where events are not too sparse, but the discriminator must extrapolate well up to the $0\nu\beta\beta$ region of interest. This requires training on a mostly clean sample of pulses and then normalizing out energy information.

\subsection{Training and Normalization}
\label{sec:TrainingNorm}

\begin{figure}
\centering
\includegraphics[width=0.46\textwidth]{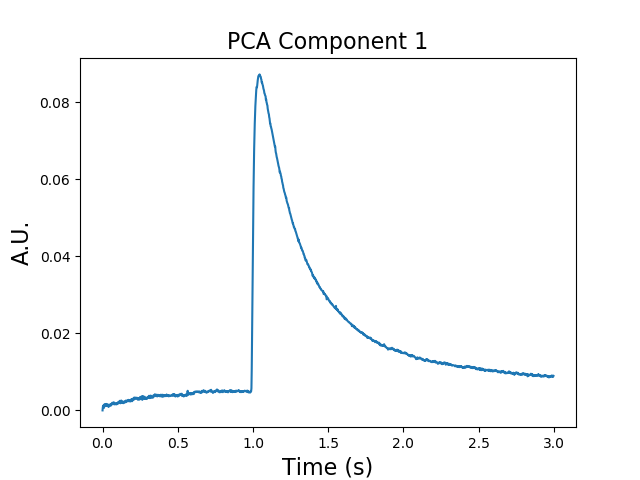}
\includegraphics[width=0.46\textwidth]{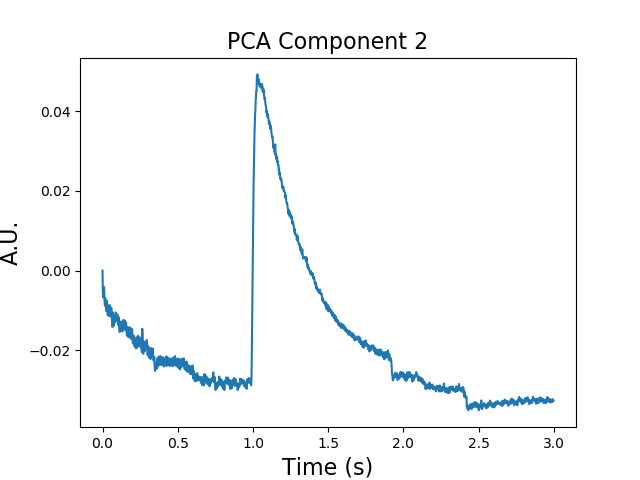}
\includegraphics[width=0.46\textwidth]{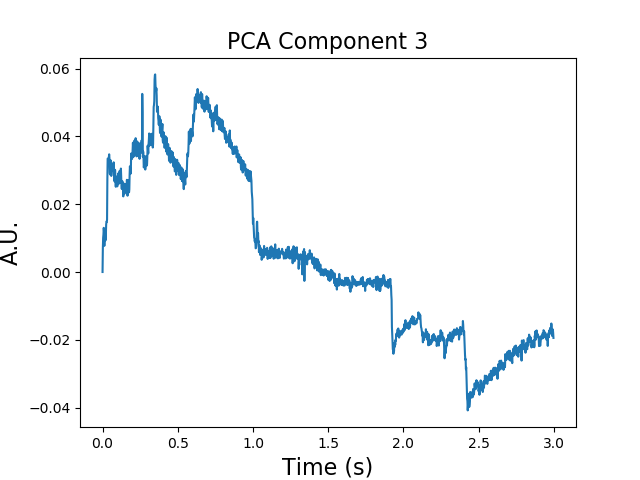}
\includegraphics[width=0.46\textwidth]{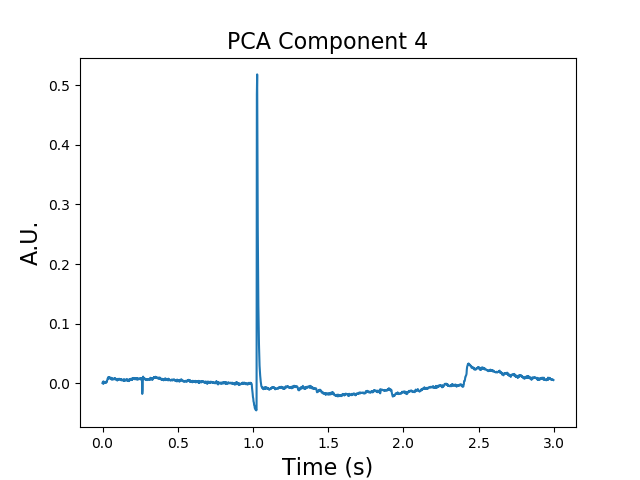}
\includegraphics[width=0.46\textwidth]{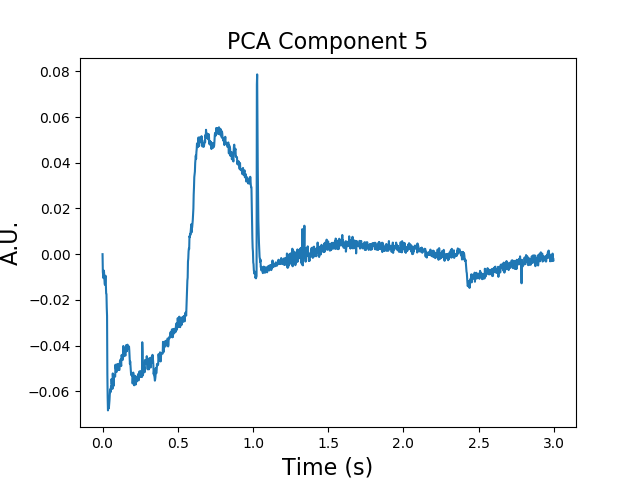}
\includegraphics[width=0.46\textwidth]{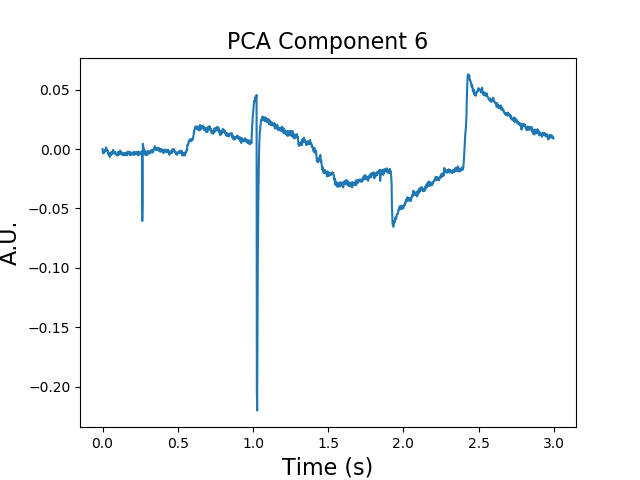}
\caption{A typical example of the 6 leading PCA components for the pulses from one LMO channel in a dataset. The first two components prominently feature the idealized pulse response of the bolometer, while the following components tend to be dominated by other features. Components 4 through 6 notably feature information about the timing of the pulse.}
\label{fig:PCALeadingComponent}
\end{figure}

To create a training sample for each LMO detector, we collect events falling under the following requirements:
\begin{enumerate}
	\item Triggered events with energy between 1 and 2 MeV in physics data (defined as data collected without a calibration source inserted into the detector).
	\item Passing a light yield cut: requiring an appropriate amount of light to be seen in the adjacent light detectors.
	\item Passing a multiplicity cut: no triggered events within a small time window in any of the other LMO detectors in the cryostat.
\end{enumerate}
The large majority of events satisfying these properties will be standard model two-neutrino double-beta decays ($2\nu\beta\beta$) of $^{100}$Mo, with an additional $<10\%$ contribution from $\gamma$ backgrounds \cite{CUPIDMo2vResult}. The physical nature of a $2\nu\beta\beta$ energy deposit looks exactly like a $0\nu\beta\beta$ event, besides the different total amount of energy, so this serves as a good signal-like sample. The light yield and event multiplicity cuts also remove most events that are caused by abnormal detector conditions, leaving mostly physical events. Since $^{100}$Mo has a relatively high $2\nu\beta\beta$ decay rate ($\sim2$ mHz per detector), this still leaves well over a thousand events per LMO channel in a typical 3 to 5 week dataset to fill the training sample. The typical leading PCA components extracted from this training procedure look like those shown in Fig. \ref{fig:PCALeadingComponent}, with the first two capturing the general shape of a ``good'' pulse. This result can be understood as a statement that the primary variation between good pulses is in their different amplitudes, corresponding to their different energies. Other localized features or pulse shape differences are identified as subleading effects, and appear in the subleading components. 

\begin{figure}
\centering
\includegraphics[width=0.85\textwidth]{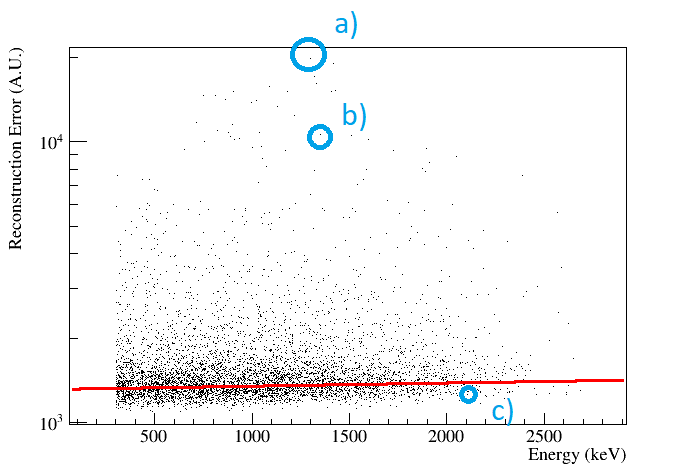} \\
\includegraphics[width=0.48\textwidth]{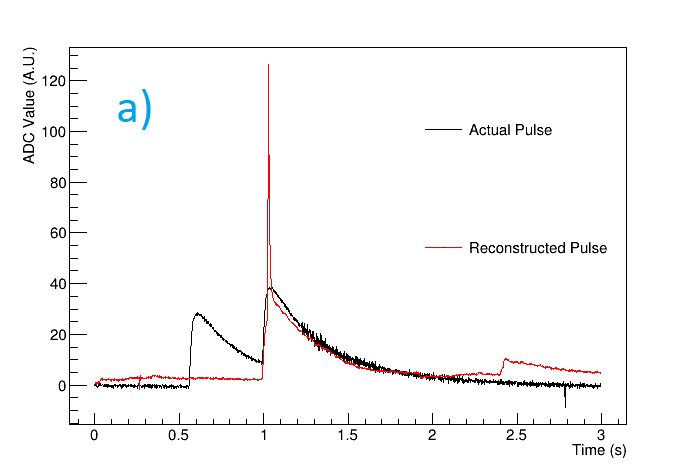}
\includegraphics[width=0.48\textwidth]{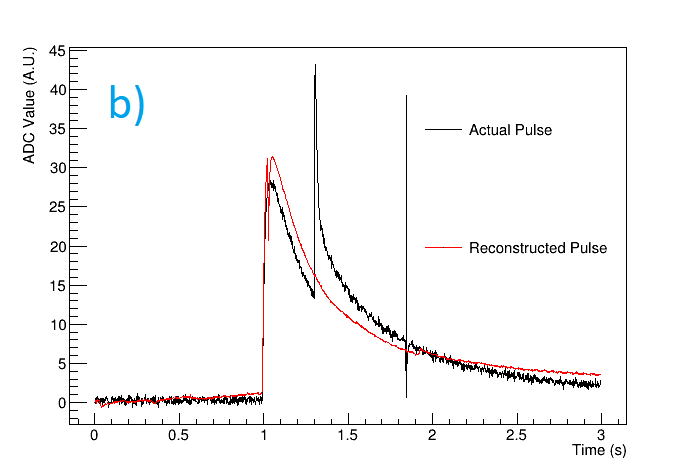} \\
\includegraphics[width=0.48\textwidth]{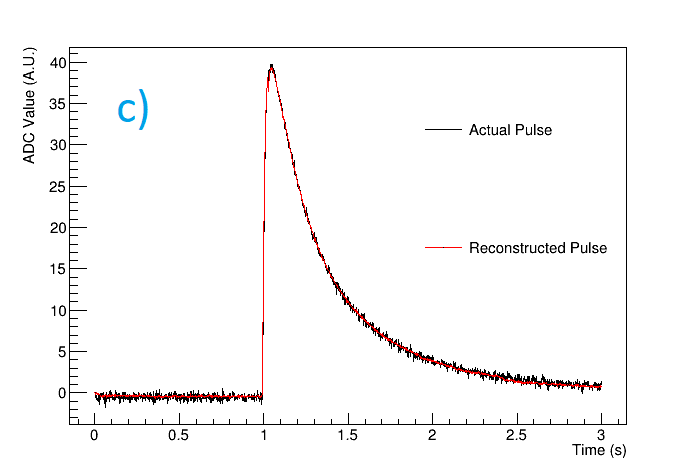}
\caption{Top: Scatterplot of event reconstruction errors versus energy for one LMO channel in a dataset, the same shown in Fig. \ref{fig:PCALeadingComponent}. The red line is the linear fit used to normalize the reconstruction errors as a function of energy, obtained using a robust method to prune outlier points from the fit. Three points a), b), and c) are marked as examples of rejected and accepted pulses. Bottom: The labeled example pulses are drawn in black, with the attempted reconstruction using the leading 4 PCA components overlaid in red. Pulses a) and b) here have large reconstruction errors compared to the base population and end up rejected as anomalous pulses, while pulse c) falls within the normal range of error and is accepted as a good pulse.}
\label{fig:PCANormFit}
\end{figure}

Applying these components to the rest of the data collected in a dataset, we expect the reconstruction error to have some mild energy dependence, since the shape of a pulse is not completely the same at all energies. This energy dependence must be calibrated out if we intend to apply a cut based on this metric. We empirically observe that the dependence is linear for all of our detectors. The reconstruction errors can then be normalized by performing a linear fit against energy for the bulk of events, using a robust fitting procedure to avoid bias from outliers. An example of one such fit is shown in Fig. \ref{fig:PCANormFit}, along with examples of pulses that are eventually accepted and rejected through this procedure. Once this fit is obtained, the reconstruction error of each event can be normalized using its deviation from the ``predicted'' reconstruction error provided by the fit at that energy. It should be noted we must assume that the good pulses comprise the majority of events selected for the fitting procedure, but this is a safe assumption for CUPID-Mo since most spurious events are removed by light yield and multiplicity cuts already.

\subsection{Results}

\begin{figure}
\centering
\includegraphics[width=0.8\textwidth]{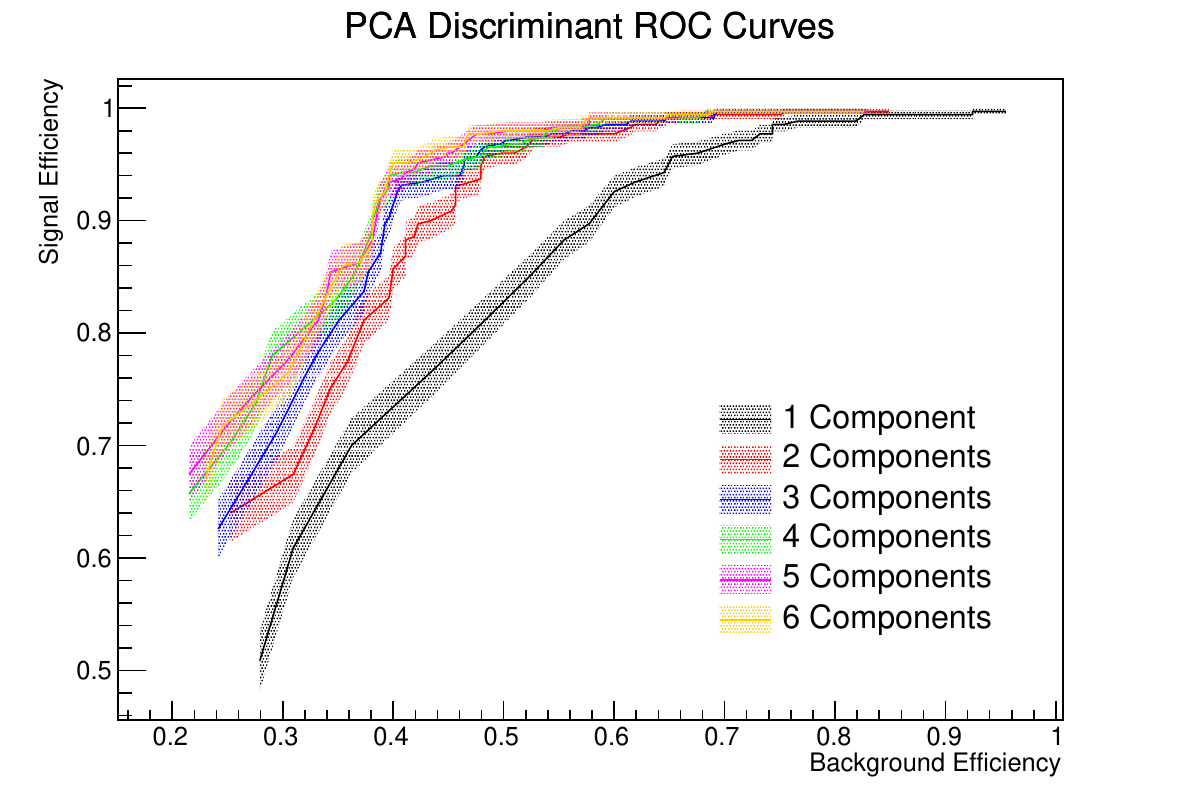}
\caption{ROC curves and associated error bands for the PCA-based discriminant with the reconstruction error calculated using varying number of components. The signal efficiency is evaluated on the 2615 keV $^{208}$Tl $\gamma$ peak, and the background efficiency is evaluated on $\beta/\gamma$-like events in the pileup region of 2750 to 3000 keV in calibration.}
\label{fig:ROCCurve}
\end{figure}

Through this procedure of calculating PCA-based reconstruction errors and performing the fitting procedure described in Sec. \ref{sec:TrainingNorm}, we obtain normalized reconstruction errors for each event. One question is how many PCA components to use in the analysis when calculating the reconstruction errors. We want to use enough components to capture signal-like behavior, but not so many that noise-like events can be adequately covered and assigned a low reconstruction error. Standard receiver operating characteristic (ROC) curves can be used to compare the signal and background efficiencies of this PCA-based discriminant for varying numbers of components, as shown in Fig. \ref{fig:ROCCurve}. Signal efficiency is evaluated on 2615 keV $^{208}$Tl $\gamma$ events, which should in general have very similar detector response characteristics to a $0\nu\beta\beta$ event. Since we already have light yield cuts that eliminate almost all $\alpha$ events around the region of interest in CUPID-Mo, much of the remaining background is expected to be pileup of $\beta/\gamma$ events. We thus evaluate background efficiency on $\beta/\gamma$ events in the 2750 to 3000 keV region in calibration data, where we have radioactive sources inserted into the cryostat. Our highest energy calibration line is from 2615 keV $^{208}$Tl $\gamma$ events, so many events in this energy region are the pileup events we want to reject. Pileup at this energy is rare in physics data, which has much lower event rates, and so we must evaluate this efficiency in calibration data in order to have sufficient statistics. However, it should be noted that our background sample for this calculation is not pure, as a non-negligible number of events in this energy region will still be signal-like events. As a result, we should not expect a near-100\% rejection of background to be possible. The ROC curves show that we obtain a notable improvement going from 1 component to 2 components, but that improvements taper off as we add even more components. We choose to use 4 components in our analysis, seeing that there is no discernible benefit from using more than that.

\begin{figure}
\centering
\includegraphics[width=0.8\textwidth]{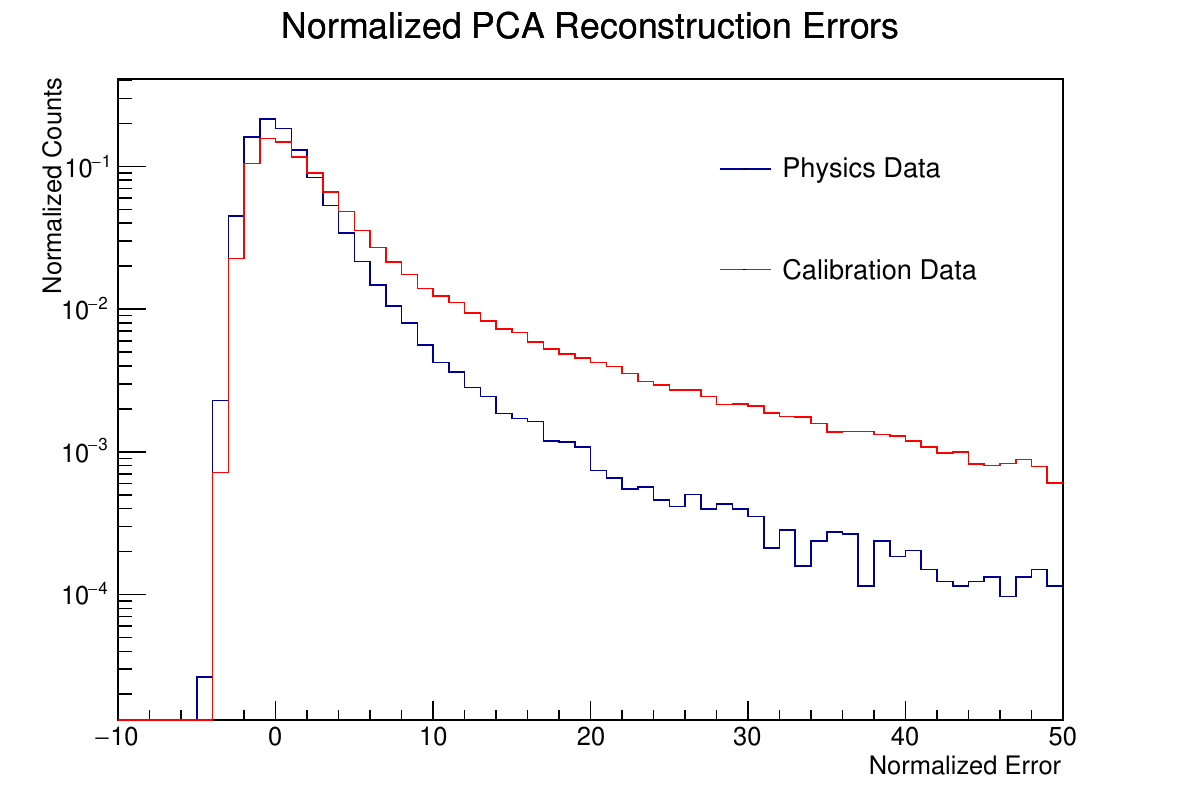}
\caption{Distribution of the normalized PCA reconstruction error variable in both calibration and physics data, after the normal data quality cuts and light yield cuts. Events passing these cuts are all $\beta / \gamma$-like, so any remaining anomalies are mostly from either pileup or from detector effects that slipped through the basic cuts. The distributions shown have their total counts normalized for comparison. It can be seen that in calibration data the PCA reconstruction error skews higher than in physics data, corresponding to the higher rate of pileup during calibration.}
\label{fig:NormErrorDistribution}
\end{figure}

\begin{figure}
\centering
\includegraphics[width=0.8\textwidth]{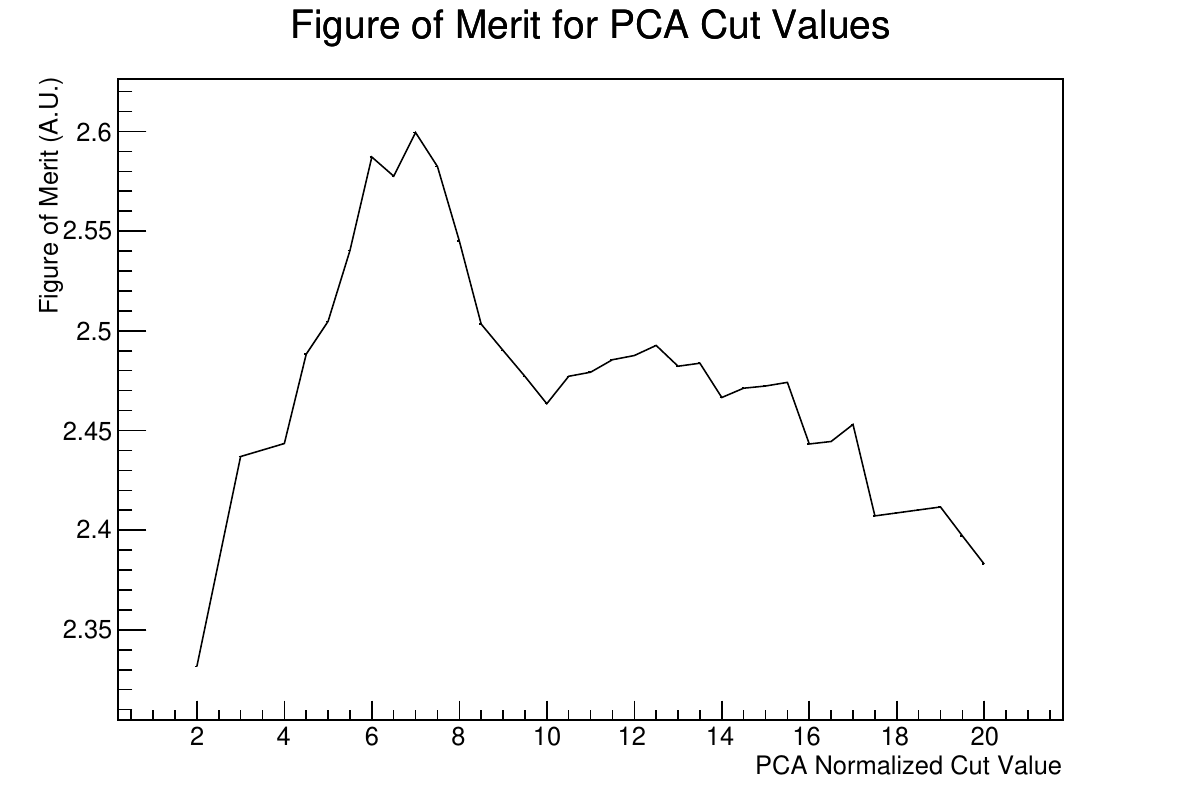}
\caption{Figure of merit $\sqrt{2((S+B)\ln{(1+S/B)}-S)}$ for various cut values on the PCA normalized error metric. Signal rate is evaluated with the previous NEMO-3 limit on the $^{100}$Mo $0\nu\beta\beta$ rate times the cut efficiency on the background $^{208}$Tl gamma peak. Background rate is evaluated as the number of calibration events in the 2880 to 2980 keV region, which are expected to be primarily due to pileup, scaled down by the $^{208}$Tl rate in physics data versus calibration data.}
\label{fig:FigureOfMerit}
\end{figure}

The distribution of this normalized error variable using the 4 leading PCA components for one of our datasets is shown in Fig. \ref{fig:NormErrorDistribution}. The shape of the distribution is as expected, with a core around 0 and then a tail extending to high values. The tail is also more prominent in calibration data, as we would expect due to the much higher occurrence of pileup events. Since we expect to have very low levels for both background and signal in the $0\nu\beta\beta$ region of interest, we select our final cut value by optimizing the median discovery significance $\sqrt{2((S+B)\ln{(1+S/B)}-S)}$ suggested by Cowan et al. in \cite{CowanMetric}. For the purposes of cut optimization, we use the previous limit on the $0\nu\beta\beta$ rate of $^{100}$Mo from NEMO-3 \cite{NEMO3Limit} for our expected signal, scaled down by the cut's signal efficiency. For our expected pileup background, we use the number of 2880-2980 keV events in calibration data that pass the PCA cut and scale it down by the $^{208}$Tl event rate in physics data compared to calibration. The resulting figure of merit for different cut values is shown in Fig. \ref{fig:FigureOfMerit}, giving an optimal cut value of 7 on the normalized error. This corresponds to a signal efficiency of $(94.9 \pm 1.2)\%$ and a background acceptance of $(41.2 \pm 2.9)\%$ on the ROC curve in Fig. \ref{fig:ROCCurve}, although it should again be noted that since the background sample contains signal-like events, this should not be considered the cut's true background rejection capability.

\begin{figure}
\centering
\includegraphics[width=0.8\textwidth]{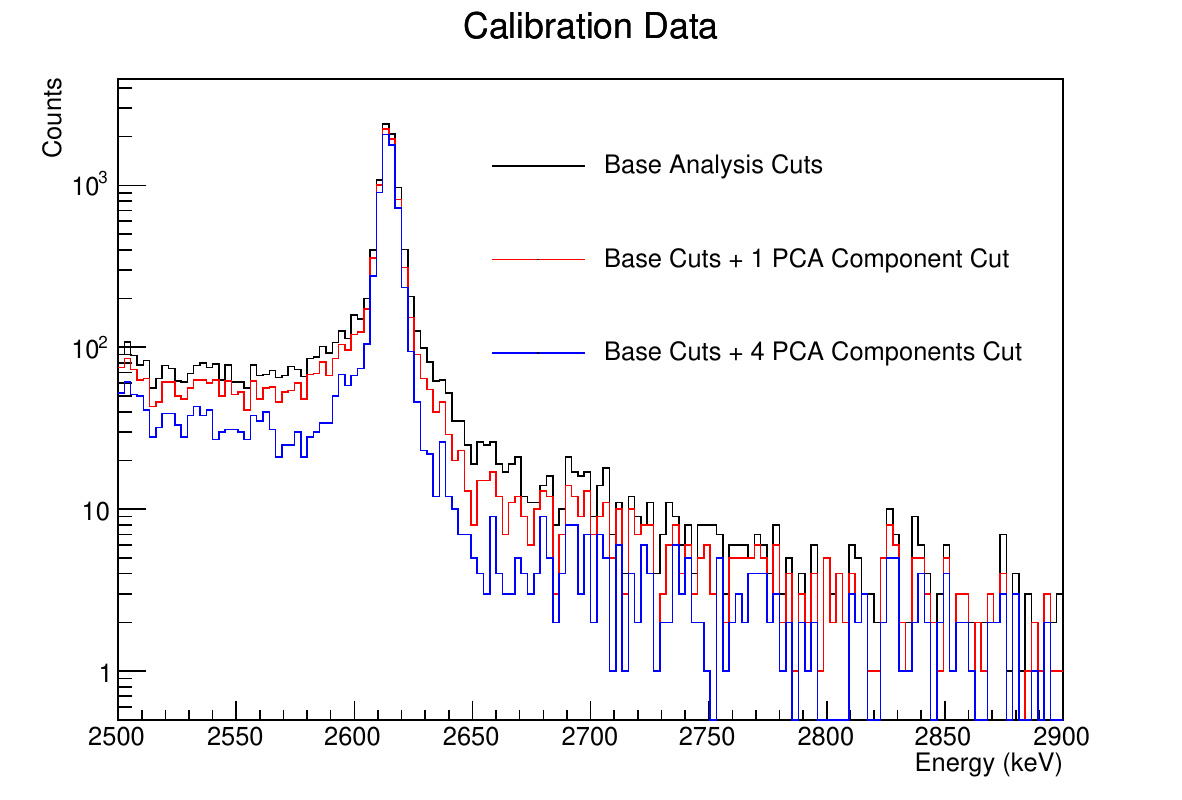}
\caption{The $^{208}$Tl $\gamma$ peak at 2615 keV from a calibration period, with the basic data quality and light yield cuts in black, with an additional cut on the 1-component PCA reconstruction error in red, and with the 4-component PCA cut in blue. We see that the PCA-based cut mostly eliminates events from the tails of the distribution, where there are more ``bad'' pulses compared to the peak at 2615 keV. We also see using 4 components instead of 1 mostly preserves events in the peak, but narrows its width and rejects more events from the sidebands.}
\label{fig:CalibrationPeak}
\end{figure}

\begin{figure}
\centering
\includegraphics[width=0.8\textwidth]{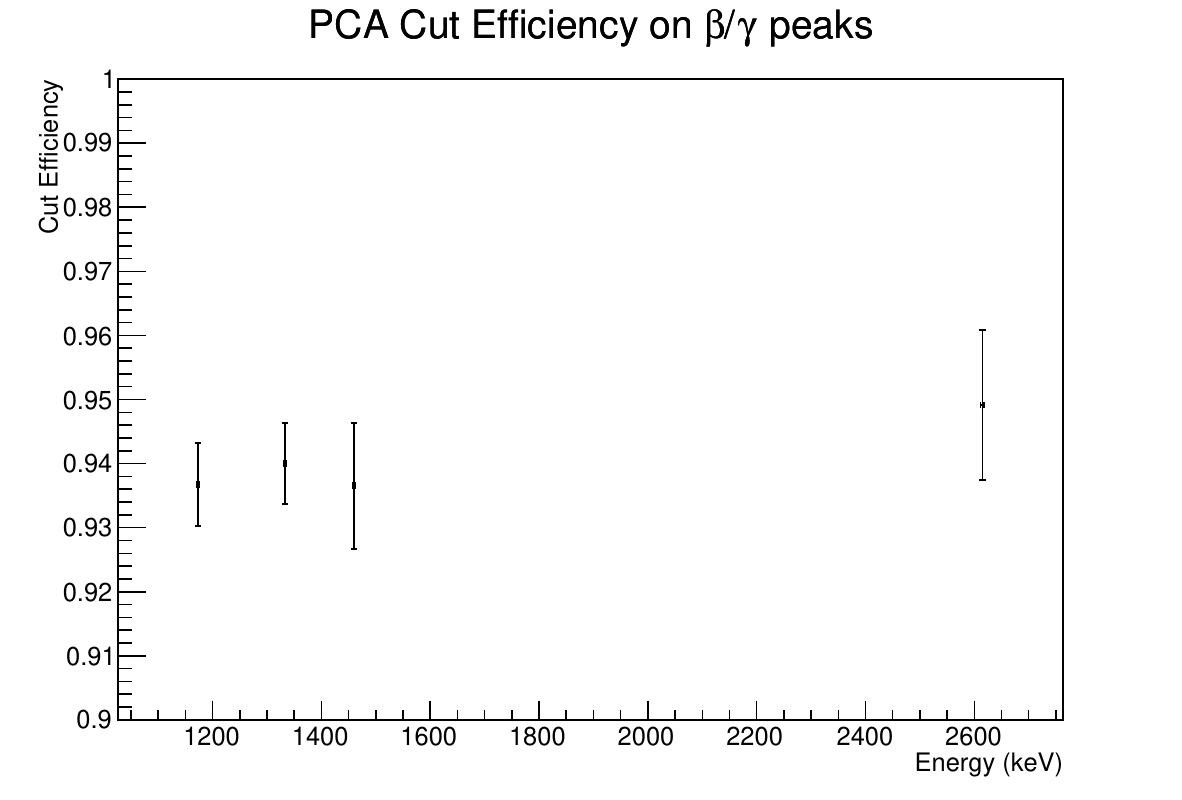}
\caption{PCA cut efficiencies evaluated on the notable $\beta/\gamma$ peaks in the background spectrum at 1173, 1333, 1460, and 2615 keV, from ambient $^{60}$Co, $^{40}$K, and $^{208}$Tl that is present even for physics data. We see no evidence of an energy dependence of the cut efficiency in this range.}
\label{fig:Efficiencies}
\end{figure}

\begin{figure}
\centering
\includegraphics[width=0.8\textwidth]{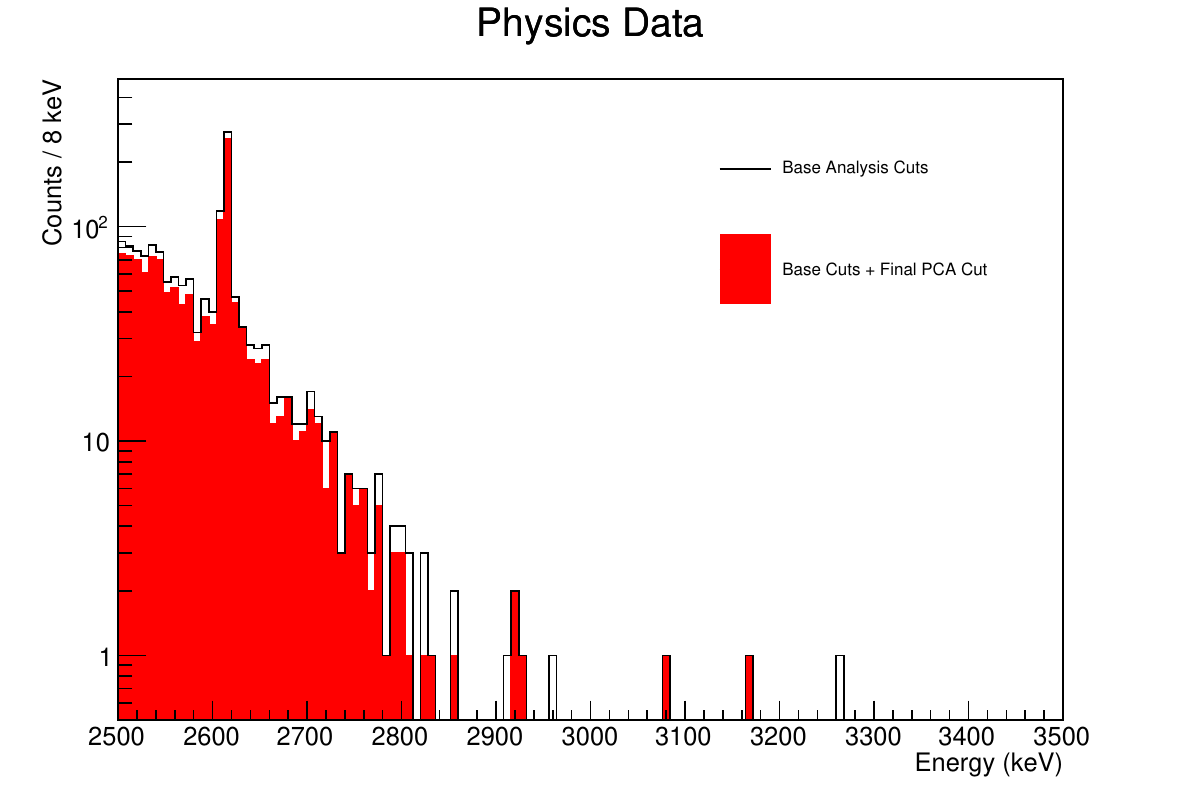}
\caption{Physics data including the $^{100}$Mo $0\nu\beta\beta$ region of interest, for a total exposure of 2.16 kg$\times$yr. Base analysis cuts include multiplicity and light yield cuts. We see the PCA cut mostly preserves events in the $\beta/\gamma$ region while cutting a significant number of the few events above 2800 keV that comprise our background for a $0\nu\beta\beta$ search.}
\label{fig:BackgroundROI}
\end{figure}

We then apply a simple binary cut using this value, optimizing our acceptance of $0\nu\beta\beta$-like events compared to rejection of expected backgrounds in the $0\nu\beta\beta$ energy region. Fig. \ref{fig:CalibrationPeak} shows that the cut based on the PCA reconstruction error mostly trims events from the tails of the calibration spectrum, which is largely populated by pileup events. At the same time, the cut mostly keeps events within the calibration peak, which should mostly be good pulses. The efficiency of this cut on signal-like events is evaluated on a number of $\beta/\gamma$ peaks in the physics data by fitting them with a Gaussian signal on top of a background and then counting the number of events that pass the cut after background subtraction, a procedure previously outlined in \cite{CUORE0Analysis}. We see that the efficiency is flat in the energy regions of interest, as shown in Fig. \ref{fig:Efficiencies}, indicating the cut successfully avoids eliminating events based primarily on energy and extrapolates well to the $0\nu\beta\beta$ Q-value. The effect of the final PCA-based pulse shape cut on the physics data for an exposure of 2.16 kg$\times$yr is shown in Fig. \ref{fig:BackgroundROI}. The $0\nu\beta\beta$ region of interest around 3034 keV already had no events even before the PCA cut, but looking at a broader energy range we can see the cut reduces the number of background events in the 2900-3500 keV region from 8 to 5.

\section{Conclusion}

With the background suppression offered by light detectors in CUPID-Mo and in the future CUPID experiment, pileup and spurious events from detector effects will become leading background components in the region of interest for $0\nu\beta\beta$ decay of $^{100}$Mo. We have demonstrated that a principal component analysis based approach is effective for filtering out these types of anomalous pulses in CUPID-Mo, even in the absence of simulations to provide clean training and testing samples. Using just a moderately clean sample of pulses pulled directly from the data, we can obtain signal-like components from the PCA decomposition, and we are able to use the corresponding reconstruction errors to construct an energy-independent metric that can be used to reject undesirable events. This method shows promise for pulse shape discrimination in cryogenic bolometric detectors and has the potential to significantly reduce pileup backgrounds in CUPID. Further tuning is likely still possible by focusing on specific time windows of the pulse, in particular around the trigger time for very close pileup events. The development of detector response simulation tools will also aid in the training and testing procedure and may enable supervised machine learning algorithms based on PCA outputs.

\section{Acknowledgements}

This work has been partially performed in the framework of the
LUMINEU program, a project funded by the Agence Nationale de la
Recherche (ANR, France). The help of the technical staff of the
Laboratoire Souterrain de Modane and of the other participant
laboratories is gratefully acknowledged. F.A. Danevich, V.V.
Kobychev, V.I. Tretyak and M.M. Zarytskyy were supported in part
by the National Research Foundation of Ukraine Grant No.
2020.02/0011. O.G. Polischuk was supported in part by the project
``Investigations of rare nuclear processes'' of the program of the
National Academy of Sciences of Ukraine ``Laboratory of young
scientists''. A.S. Barabash, S.I. Konovalov, I.M. Makarov, V.N.
Shlegel and V.I. Umatov were supported by Russian Science
Foundation (grant No.  18-12-00003). The Ph.D. fellowship of H.
Khalife has been partially funded by the P2IO LabEx
(ANR-10-LABX-0038) managed by the ANR (France) in the framework of
the 2017 P2IO Doctoral call. We acknowledge the support of the P2IO LabEx (ANR-10-LABX-0038) in the framework ``Investissements d’Avenir'' (ANR-11-IDEX-0003-01 – Project ``BSM-nu'') managed by the Agence Nationale de la Recherche (ANR), France. C. Rusconi is supported by the
National Science Foundation Grant NSF-PHY-1614611. This material
is also based upon work supported by the US Department of Energy
(DOE) Office of Science under Contract No. DE-AC02-05CH11231; by
the DOE Office of Science, Office of Nuclear Physics under
Contract Nos. DE-FG02-08ER41551 and DE-SC0011091; by the
France-Berkeley Fund, the MISTI-France fund, and by the
Chateaubriand Fellowship of the Office for Science \& Technology
of the Embassy of France in the United States.
This work makes use of the DIANA data analysis software which has been
developed by the Cuoricino, CUORE, LUCIFER, and CUPID-0
Collaborations. 

\bibliographystyle{JHEP}
\bibliography{Bibliography}

\end{document}